\documentclass[a4paper]{article}
\usepackage{amsfonts}
\usepackage{amssymb}
\usepackage[margin=2cm]{geometry}
\usepackage[dvips]{graphicx}

\usepackage[T1]{fontenc}
\usepackage[latin2]{inputenc}

\usepackage{amsmath}
\usepackage{bbm}

\begin{document}


\title{Quantumness of Josephson junctions reexamined. }

\author{Robert Alicki \\ 
  {\small
Institute of Theoretical Physics and Astrophysics, University
of Gda\'nsk,  Wita Stwosza 57, PL 80-952 Gda\'nsk, Poland}\\
}

\date{\today}
\maketitle

\begin{abstract}
There exists an increasing evidence supporting the picture of the Josephson junction (JJ)  as a "macroscopic quantum system". On the other hand the interpretation of experimental data strongly depends on the assumed theoretical model.  We analyse the possible states of a Cooper pair box ("charge qubit") for the two types of models : two-mode Bose-Hubbard model with its large $N$ aproximations  and the many-body description within the mean-field approximation (Gross-Pitaevski equation). While the first class of models  supports the picture of JJ being a quantum subsystem of a single degree of freedom, the second approach yields an essentially classical structure of accessible quantum states which, in particular, implies the absence of  entanglement for two coupled JJ's. The arguments in favor of the mean-field theory are presented and different experimental tests including a new proposal are briefly discussed.

\end{abstract}
\section{Introduction}

The controversy concerning the classical or quantum character of the Josephson junction (JJ) has a long history. In the recent years the picture of JJ as a "macroscopic quantum system"\cite{CL} prevails and even more, JJ technology is considered to be a very promissing implementation of quantum information processing \cite{D},\cite{W}. The experimental demonstration of  "quantum superpositions" of states for a single JJ \cite{N}, spectroscopic evidence of coupling between two such devices \cite{B} and finally the very recent "measurement of the entanglement of two superconducting qubits via state tomography" \cite{MS} seem to strongly support this point of view. However, some new theoretical arguments in favor of the opposite view - "JJ is essentially a classical system"- appeared also \cite{RA}.
\par
The aim of this paper is to analyse carefully the different mathematical models of JJ using the so-called \emph{Cooper pair box} (CPB) as a concrete example. We begin with the two-mode Bose-Hubbard model and show its equivalence, for a large number of Cooper pairs $N$, to another 
one - the quantum-phase model with proper boundary conditions. These two models essentially support the "naive" picture of quantisation
of a nonlinear superconducting circuit which leads to a quantum nonlinear oscillator (quantum pendulum model) \cite{A}. Lowest lying two energy eigenstates of such a system yield a qubit in the standard approach.
\par
Then a mean-field many-body theory based on the Gross-Pitaevski equation is applied to a single CPB and two coupled CPB's. 
The obtained structure of the accessible many-particle states is shown to be completely different from the predicted by the previous theories. A single CPB cannot support a qubit and the coupling between two CPB's introduces classical correlations instead of quantum entanglement. Strictly speaking the amount of entanglement can be of the order of $1/N$, only. 
\par
To compare both approaches a new Hamiltonian obtained as a large $N$ limit of the two-mode Bose-Hubbard model is derived. This Hamiltonian describes a certain nonlinear oscillator model with coherent vectors in a deep semiclassical regime directly corresponding to quantum states of the mean-field model. It is argued that only those coherents states can be observed in experiments in contrast to unstable eigenvectors of the Hamiltonian.
\par
Finally, the alternative interpretations of a number of existing experimental tests and the proposal for a new one are briefly discussed.

\section{The theories of Cooper pair box}
We consider a simplified model of  JJ assuming that  Cooper pairs can be treated as a bosonic gas
below the critical temperature of Bose-Einstein condensation. We have two electrodes  made of a superconducting material separated by a thin layer of an insulator which allows for tunneling of Cooper pairs. In order to construct a superconducting qubit one tries to suppress the tunelling of many Cooper pairs by using the circuit consisting of a small superconducting island "1" connected  via JJ to a large superconducting reservoir "2". Coulomb
interaction between Cooper pairs in a small electrode become important and must be taken into account in the Hamiltonian.
\subsection{Two-mode Bose-Hubbard model}
The annihilation and creation operators  ${\hat a}_1, {\hat a}_1^{\dagger}$ and ${\hat a}_2 , {\hat a}_2^{\dagger}$  correspond to the
ground states of a boson (Cooper pair) in separated electrodes "1" and "2", respectively.
Coulomb interaction between Cooper pairs in a small electrode is modelled by  the quadratic term  in the Hamiltonian below. The second term contains the potentials of the both electrodes and the term proportional to $K$ describes the tunneling
of Cooper pairs
\begin{equation}
{\hat H}_{2mod} = E_C( {\hat a}_1^{\dagger}{\hat a}_1)^2 +\frac{1}{2} (U_1{\hat a}_1^{\dagger} {\hat a}_1 + U_2 {\hat a}_2^{\dagger}{\hat a}_2)  - K ({\hat a}_1 {\hat a}_2^{\dagger} + {\hat a}_1^{\dagger}{\hat a}_2)\ .
\label{2mod}
\end{equation}
The total number of bosons if fixed and equal to $N$, hence we should consider the restriction ${\hat H'}_{2mod}$ of the Hamiltonian (\ref{2mod}) to the $N+1$-dimensional Hilbert subspace of physical states spanned by the orthonormal basis 
\begin{equation}
|k\rangle = [k!(N-k)!]^{-1/2} \bigl({\hat a}_1^{\dagger}\bigr)^k({\hat a}_2^{\dagger}\bigr)^{(N-k)}|vac\rangle\ .
\label{bas}
\end{equation}
${\hat H'}_{2mod}$ is defined as follows (we omit an irrelevant constant and use the convention $|N+1\rangle=|-1\rangle=0$)
\begin{equation}
{\hat H'}_{2mod}|k\rangle = \bigl[E_C (k-{\bar n}_1)^2\bigr]|k\rangle - K\sqrt{k(N-k)}(|k+1\rangle + |k-1\rangle)\ ,\ k=0,1,...,N.
\label{2mod'}
\end{equation}
where $U = U_2 - U_1$ and ${\bar n}_1 = U/4E_C$ is an averaged number of Cooper pairs on the island corresponding to a stationary reference state. As we are interested in the low lying energy eigenstates we can replace in the second term $K\sqrt{k(N-k)}$ by $K\sqrt{{\bar n}_1(N-{\bar n}_1)}\equiv E_J$. Therefore, the final form, which can be called the Bose-Hubbard Hamiltonian, reads
\begin{equation}
{\hat H}_{B-H}|k\rangle = \bigl[E_C (k-{\bar n}_1)^2\bigr]|k\rangle - E_J(|k+1\rangle + |k-1\rangle)\ ,\ k=0,1,...,N.
\label{B-H}
\end{equation}
with the convention $|N+1\rangle=|0\rangle,|-1\rangle= |N\rangle$. This Hamiltonian can be interpreted as describing a particle with a discretized position $k$ in a harmonic potential located around $k={\bar n}_1$
and a hopping term modelling its free motion.
\par

\subsection{From two-mode to quantum phase model}
The standard picture of the CPB can be obtained from the Hamiltonian (\ref{2mod}) treating the operators
${\hat a}_1, {\hat a}_2$ as classical variables $a_1 , a_2$ and using the following parametrization 
\begin{equation}
n_1 = a_1^*a_1,\ n_2=a_2^*a_2, \ a_1= \sqrt{n_1}e^{i\theta_1},\ a_2= \sqrt{n_2}e^{i\theta_2} ,\ \theta_2 -\theta_1=\theta,\ n_1 + n_2 = N,
\label{cl}
\end{equation}
which, under the condition
$|n_1- {\bar n}_1| << {\bar n}_1 $, yields 
the Hamiltonian (up to an irrelevant constant)  
\begin{equation}
H = E_C(n_1- {\bar n}_1 )^2 - E_J \cos\theta\ .
\label{phmod}
\end{equation}
The next step in this rather "naive" approach is to "requantize" the variables $n_1 , \theta$ assuming the following
canonical commutation relations 
\begin{equation}
[{\hat n}_1 , {\hat \theta}]=i \ ,
\label{ccr}
\end{equation}
and to consider the properties of the "quantum nonlinear oscillator" with the Hamiltonian
\begin{equation}
{\hat H} = E_C{\hat\xi}^2 - E_J \cos{\hat\theta}\ ,
\label{phmodq}
\end{equation} 
where ${\hat\xi}={\hat n}_1-{\bar n}_1$ is a "momentum" operator conjugate to ${\hat\theta}$. The two lowest lying eigenstates of ${\hat H}$, which are automatically orthogonal, are supposed to provide the superconducting implementation of a qubit.
\par
A more subtle and mathematically sound approach starts with the Hamiltonian (\ref{B-H})
and treats the basis $|k\rangle$ as a subset $|k\rangle \equiv (1/\sqrt{2\pi}) e^{i(k-{\bar n}_1)\theta} ; k=0,1,...N$
of the (rotated by $e^{-i{\bar n}_1\theta}$) Fourier basis in the quantum rotator  Hilbert space $L^2[-\pi,\pi]$. Assume that ${\bar n}_1$ is a natural number. Then on the subspace of wave functions $f(\theta)$ spanned by these particular basis elements
\begin{equation}
f(\theta) = \sum_{k=0}^N c_k e^{i(k-{\bar n}_1)\theta}=\sum_{k=-{\bar n}_1}^{N-{\bar n}_1} c_{k+{\bar n}_1} e^{ik\theta}\ , c_k \in {\bf C}
\label{basis}
\end{equation} 
the action of the operator ${\hat H}_{B-H}$  is equivalent to the action of the following differential operator
\begin{equation}
{\hat H}_{qph}f(\theta) = -E_C \frac{\partial^2}{\partial\theta^2}f(\theta) -
E_J \cos \theta f(\theta) 
\label{qph}
\end{equation}
with the periodic boundary condition $f(\pi) = f(-\pi)$.
Then we treat the operator ${\hat  H}_{qph}$ extended to the whole Hilbert space $L^2[-\pi,\pi]$ as a good approximation to the original Hamiltonian (\ref{B-H}). This is, indeed, a reasonable procedure because $N >>{\bar n}_1$ and ${\bar n}_1$
is at least of the order of $10^8$. Hence, the restriction to a finite sum in (\ref{basis}) becomes irrelevant at least for the interesting low energy regime.  We can continuously change the potential difference $U$ to produce a continuous change of the parameter ${\bar n}_1 = U/4E_C$ beyond the natural numbers. This disagrees  with the necessary periodicity of the functions (\ref{basis}) from the domain of the Hamiltonian ${\hat  H}_{qph}$. Such a generic situation leads to the slightly modified Hamiltonian $(a= {\bar n}_1- integer\ part[{\bar n}_1])$
\begin{equation}
{\hat H}_{qph}f(\theta) = -E_C \Bigl[\frac{\partial}{\partial\theta}-ia\Bigr]^2 f(\theta) -
E_J \cos \theta f(\theta)\ . 
\label{qph1}
\end{equation}

\par
\subsection{Mean-field many particle model}

The fundamental property of the Bose-Einstein condensate is the form of its time-dependent N-boson wave function $\Psi(t)$ (i.e. the true quantum state of the system) which is approximately given by a tensor product of the identical single-boson wave functions $\psi(t)$ \cite{KH}
\begin{equation}
\Psi(t) = \bigotimes_{N} \psi(t) \ . 
\label{prod}
\end{equation}
One should notice that the set of condensate's wave functions does not form a linear subspace in the Hilbert space of the N-particle system.
\par
There are numerous arguments, including theorems in mathematical physics \cite{L},\cite{E} that for weakly interacting Bose gas the initial product state evolves approximately into the product state (\ref{prod})
such that the single-particle wave function $\psi(t)$ satisfies the Gross-Pitaevski equation which is a nonlinear Schr\"odinger equation with a cubic term modelling boson-boson interaction \cite{GP}. It implies immediately that the wave function $\psi(t)$ does not correspond to a quantum state of any subsystem and the superposition principle for $\psi(t)$ is valid only approximatively, due to nonlinearity. Hence, $\psi(t)$ is  rather  a classical object: the \emph {order parameter} labelling different thermodynamical phases of the system below the critical temperature. 
\par
For our model the spatial dependence of the wave function $\psi$ is irrelevant and therefore we represent $\psi$  by the two complex amplitudes $\psi = (\psi_1 ,\psi_2)$ corresponding to the electrodes "1" and "2", respectively. They satisfy a discrete version of the Gross-Pitaevski equation which,  for the convenience, is formulated in terms of the renormalized wave function $\phi = \sqrt{N}\psi$ \cite{S}
\begin{eqnarray}
\dot{\phi}_1 =& -i (U_1 + g |\phi_1|^2)\phi_1  -iK\phi_2
\nonumber\\
\dot{\phi}_2 =& -i U_2 \phi_2  -iK\phi_1
\label{GP} 
\end{eqnarray}
with $|\phi_1|^2 +|\phi_2|^2 =N$. The term proportional to $g$ describes Coulomb interaction of Cooper pairs in the first electrode in the mean-field approximation.
\par
Introducing the following parametrization 
\begin{equation}
\phi_j = q_j +i p_j= \sqrt{n_j}e^{i\theta_j}\ ,\ j=1,2
\label{par}
\end{equation}
one can easily check that the equations
(\ref{GP}) can be written as the Hamiltonian equations for the system of 2-degrees of freedom
governed by the classical Hamiltonian
\begin{equation}
h_2(q_1,q_2,p_1,p_2) = \frac{U_1 -U_2}{2}(p_1^2 +q_1^2) + \frac{g}{4}(p_1^2 +q_1^2)^2 - K(p_1p_2 +q_1 q_2) +
\frac{U_2}{2}(p_1^2 +q_1^2 + p_2^2 +q_2^2)\ . 
\label{cla}
\end{equation}
The system of above is integrable because of the existence of the constant of motion
$\|\phi\|^2 =(p_1^2 +q_1^2 +p_2^2 +q_2^2)=N$. We can restrict ourselves to a decoupled, single degree
of freedom subsystem parametrized by $\theta = \theta_2 -\theta_1$ and $n_1$ (\ref{par})
with the Hamiltonian function 
\begin{equation}
h_1(\theta,n_1) = \frac{g}{4}(n_1 - \frac{U}{g})^2 - K\sqrt{n_1(N-n_1)}\cos\theta 
\label{clas}
\end{equation}
where $U = U_2-U_1$. Again, in the regime $|n_1 -{\bar n}_1|<< {\bar n}_1$ the Hamiltonian (\ref{clas})
coincides with the semiclassical limit (\ref{phmod}) of the 2-mode Hamiltonian (\ref{2mod}) if
$g/4= E_C$, ${\bar n}_1 =U/g$ and $K\sqrt{{\bar n}_1(N-{\bar n}_1)}= E_J$. We can solve the corresponding Hamiltonian
equation to obtain a trajectory  $(\theta(t), \xi(t))$ with the "momentum" $\xi(t) = n_1(t)-{\bar n}_1$.
Taking into account that $\xi(t)$ is at most of the order $\sqrt{E_J/E_C}$ and eliminating the irrelevant common phase factor we can use the following approximation for $\phi = (\phi_1, \phi_2)$ 
\begin{eqnarray}
\phi_1(t) =& \sqrt{{\bar n}_1 +\xi(t)}e^{-i\theta(t)} \simeq e^{-i\theta(t)}\Bigl[ \sqrt{{\bar n}_1} +\xi(t)\Bigl(2\sqrt{(N-{\bar n}_1){\bar n}_1}\Bigr)^{-1}\sqrt{N-{\bar n}_1}\Bigr]
\nonumber\\
\phi_2(t) =& \sqrt{N-{\bar n}_1 -\xi(t)} \simeq \Bigl[ \sqrt{N-{\bar n}_1} -\xi(t)\Bigl(2\sqrt{(N-{\bar n}_1){\bar n}_1}\Bigr)^{-1}\sqrt{{\bar n}_1}\Bigr]\ .
\label{psi}
\end{eqnarray}
Now, the normalized to one single boson wave function $\psi= N^{-1/2}\phi$ can be written as the superposition of two orthonormal vectors rotated by the unitary $W$
\begin{equation}
\psi(t) = W(\theta(t))\bigl[\chi_0 +  \xi(t)\Bigl(2\sqrt{(N-{\bar n}_1){\bar n}_1}\Bigr)^{-1}\chi_1\bigr]
\label{psi1}
\end{equation}
where
\begin{equation}
W(\theta) = {\rm diag} (e^{-i\theta} , 1)\ ,\ \chi_0= N^{-1/2}\Bigl(\sqrt{{\bar n}_1},\sqrt{N-{\bar n}_1}\Bigl),\  \chi_1 = N^{-1/2}\Bigl(\sqrt{N-{\bar n}_1}, -\sqrt{{\bar n}_1}\Bigr)\ .
\label{psi2}
\end{equation}
One should notice that the amplitude of the vector $\chi_1$ is scaled down by the factor $(2\sqrt{(N-{\bar n}_1){\bar n}_1})^{-1}<<1$ and hence the normalization of $\psi (t)$ is preserved up to the higher order terms.
\par
We can reconstruct now the quantum state $\Psi(t)$ describing the condensate in the mean-field approximation keeping the lowest order correction of the order $1/\sqrt{{\bar n}_1}$
\begin{equation}
\Psi(t) = {\bf W}(\theta(t))\Bigl[\Phi_0 +  \xi(t)\sqrt{N}\Bigl(2\sqrt{(N-{\bar n}_1){\bar n}_1}\Bigr)^{-1}\Phi_1\Bigr]
\label{Psi}
\end{equation}
where
\begin{equation}
{\bf W}(\theta)=\bigotimes_N W(\theta)\ ,\ \Phi_0 =\bigotimes_N \chi_0\ ,\ \Phi_1 = \frac{1}{\sqrt{N}}\sum_{j=1}^N
\chi_0\otimes...\otimes\chi_0\otimes\underbrace{\chi_1}_{j-th}\otimes\chi_0\otimes...\otimes\chi_0\ .
\label{Psi1}
\end{equation}
One should notice that the state $\Phi_1$ does not describe the condensate but the states
of the form $\Phi_0 + \epsilon \Phi_1$ do, up to the order $\sim\epsilon^2$. It is rather clear that the obtained  2-dimensional manifold of the available states does not contain the manifold spanned by the qubit states of the form  $\alpha |0\rangle + \beta |1\rangle$ with $ |\alpha|^2+  |\beta|^2=1$. Another equivalent interpretation of  the product states (\ref{prod}) in terms of coherent vectors will be presented in Section 4.

\section{Phenomenology of the mean-field model}
The experiments performed on the different types of "JJ qubits" are usually analysed in terms of the standard quantum-phase model. The question arises whether the phenomenology of JJ can be also consistently explained within the many-particle mean-field model.

\subsection{Single Cooper pair box}

The quantum state of the controlled single CPB is described by the formula (\ref{Psi})(\ref{Psi1}) with $\theta(t),\xi(t)$ being the solutions of the classical Hamiltonian equations obtained for the Hamiltonian (\ref{clas}) with a generally time-dependent 
$U(t)$ given by
\begin{equation}
U(t) = U + u(t)\ .
\label{con}
\end{equation}
Here $U$ is a constant potential including the constant external voltage and the interaction of  Cooper pairs with the background positive charges. The term $u(t)$ describes external control usually in the form of pulses or periodic perturbations. Therefore the effective classical Hamiltonian
can be written as
\begin{equation}
h(\theta,\xi;t) = E_C \xi^2 - u(t)\xi - E_J\cos\theta \ .
\label{hamcc}
\end{equation}
Starting with the stationary state $\Psi(0,0)$ and applying a relatively weak pulse $u(t)$ we can excite the small oscillations of $\theta(t)$ and $\xi(t)$ with the plasma frequency $\omega_0\simeq \sqrt{2E_CE_J}$. From the form of the quantum state (\ref{psi}) it follows that the excess charge on the electrode "1" is equal to $2e\xi(t)$ and hence oscillates with the same frequency emitting an observable radiation. A stronger pulse increases the " kinetic energy"
$E_C \xi^2$ above the "potential barier" $2E_J$ leading to a different regime of the dynamics. Applying a periodic field one can create a variety of  resonanse phenomena and even chaotic behavior.

\subsection{Two coupled CPB's}

The coupling of two CPB's can be realized by the electromagnetic interaction between
the oscillating charges of the electrode "1" of the first CPB and the electrode "1'" of the second CPB which in the Gross-Pitaevski equations of below is described by the terms proportional to $G$.  
The coupled equations for the two wave functions of the condensate $\phi, \phi'$ read
\begin{eqnarray}
\dot{\phi}_1 =& -i (U_1 + g |\phi_1|^2 +G|\phi'_1|^2)\phi_1  -iK\phi_2
\nonumber\\
\dot{\phi}_2 =& -i U_2 \phi_2  -iK\phi_1
\nonumber\\
\dot{\phi'}_1 =& -i (U'_1 + g' |\phi'_1|^2 + G|\phi_1|^2)\phi'_1  -iK'\phi'_2
\nonumber\\
\dot{\phi'}_2 =& -i U'_2 \phi'_2  -iK'\phi'_1\ .
\label{GP2} 
\end{eqnarray}
By the procedure, completely analogical to a single CPB case, one can obtain the effective Hamiltonian for two degrees of freedom described by two angles $\theta,\theta'$ and momenta
$\xi,\xi'$
\begin{equation}
h(\theta,\theta',\xi,\xi') = E_C \xi^2 - E_J\cos\theta +E'_C \xi'^2 - E'_J\cos\theta'+\frac{G}{2}\xi\xi'\ .
\label{hamc2}
\end{equation}
Here, analogically to a single CPB case: $U = U_2-U_1$ , $U' = U'_2-U'_1$, $\xi = n_1 - {\bar n}_1$, $\xi' = n'_1 - {\bar n}'_1$, $E_C = g/4$,
$E'_C = g'/4$, $E_J = K\sqrt{(N-{\bar n}_1){\bar n}_1}$, $E'_J = K'\sqrt{(N'-{\bar n}'_1){\bar n}'_1}$ and ${\bar n}_1 ,{\bar n}'_1$ are solutions of the equations $U = g{\bar n}_1 + G{\bar n}'_1 , U'= g'{\bar n}'_1 + G {\bar n}_1$.
\par
Solving the Hamiltonian equations for the coupled nonlinear oscillators with possible external control we obtain a classical trajectory  $\{\theta(t),\theta'(t),\xi(t),\xi'(t)\}$ which defines
the evolution of a quantum state of the coupled two CPB's. This state is a tensor product of quantum states for separate CPB's, each of them given by (\ref{Psi})
\begin{equation}
\Psi(\theta,\theta', \xi,\xi') = \Psi(\theta, \xi)\otimes\Psi(\theta', \xi')\ .
\label{MF2}
\end{equation}
Therefore, it is not possible to produce entangled states within this model. A small amount of entanglement of the order $1/N$, which is allowed according to \cite{RA}, is well-placed within the accuracy limits of the mean-field approach. Obviously, the quantum states of the two JJ's are classically correlated by the coupling of the classical parameters $\theta,\xi$ and $\theta',\xi'$.

\section{Comparison of the models}

We have obtained two physically different pictures of the CPB. According to the first one
we can observe and manipulate lowest lying states and their superpositions for the  well-defined quantum system equivalent to the quantum pendulum. The second one implies that the accessible states form a nonlinear manifold parametrized by the essentially classical variables satisfying classical
equations of motion. To compare both approaches we represent the product state (\ref{prod}) in terms of the basis $|k\rangle$ introduced in (\ref{bas})
\begin{equation}
|\Psi\rangle = \Biggl(e^{-i\theta}\sqrt{\frac{n_1}{N}}\,{\hat a}_1^{\dagger}+ \sqrt{1-\frac{n_1}{N}}\,{\hat a}_2^{\dagger}\Biggr)^N|vac\rangle\ = \sum_{k=0}^N
\Biggl[\binom{N}{k} \Big(\frac{n_1}{N}\Bigr)^k\Bigl(1-\frac{n_1}{N}\Bigr)^{N-k}\Biggr]^{\frac{1}{2}}
e^{-ik\theta} |k\rangle \ .
\label{probas}
\end{equation}
Replacing now the square root of the binomial distribution in (\ref{probas}) by the square root of the Poisson one (what is consistent with the assumption $N>>n_1$) we can treat the states $|k\rangle$
as the eigenstates of ${\hat b}^{\dagger}{\hat b}$ for a certain fictitious nonlinear oscillator and the product state (\ref{probas}) becomes  the coherent vector
\begin{equation}
|\Psi\rangle \simeq  \sum_{k=0}^{\infty}\frac{\sqrt{n_1}^k}{\sqrt{k!}}e^{-\frac{n_1}{2}}e^{-ik\theta}
 |k\rangle \equiv |\sqrt{n_1}e^{-i\theta}\rangle
\label{coh}
\end{equation}
where
\begin{equation}
{\hat b} |\sqrt{n_1}e^{-i\theta}\rangle= \sqrt{n_1}e^{-i\theta}|\sqrt{n_1}e^{-i\theta}\rangle\ .
\label{coh1}
\end{equation}
Using now the identities
\begin{equation}
{\hat a}^{\dagger}_1 {\hat a}_2 |k\rangle = \sqrt{N-k}\,{\hat b}^{\dagger} |k\rangle\ ,\  
{\hat a}^{\dagger}_2 {\hat a}_1 |k\rangle = \sqrt{N-k+1}\,{\hat b}|k\rangle\
\label{ide}
\end{equation}
and the fact that the relevant states are spanned by the vectors $|k\rangle$ with $k \simeq {\bar n}_1$ we can  approximate the Hamiltonian (\ref{2mod}) by a new one
\begin{equation}
{\hat H}_{2mod}\simeq {\hat H}_{b} = E_C( {\hat b}^{\dagger}{\hat b}-{\bar n}_1 )^2   - \frac{E_J}{\sqrt{\bar n}_1} ({\hat b}+ {\hat b}^{\dagger} )\ .
\label{2mod1}
\end{equation}
\par

Here again we obtain the Hamiltonian which in the semiclassical limit yields the same equations of motion as (\ref{phmod}). In contrast to the previous approximative Hamiltonians (\ref{B-H})(\ref{phmodq}), where the large values of the quantum number $k$ are hidden by applying the shift $k\mapsto k-{\bar n}_1$, here it is clear that the lowest eigenstates of ${\hat H}_{b}$ are placed in a deep semiclassical regime corresponding to the quantum numbers (eigenvalues of ${\hat b}^{\dagger}{\hat b}$) of the order of ${\bar n}_1\simeq 10^8$. Therefore, we should not expect to observe in experiments the true eigenstates of the Hamiltonian ${\hat H}_{b}$ but rather localized "classical" states described by the coherent vectors (\ref{coh}) evolving (approximately) according to the classical Hamiltonian equations (see the next Section). 
The striking difference between the Hamiltonian eigenstates and the coherent vectors concerns the fluctuations of the Cooper pair number $n_1$ and the phase difference $\theta$. For the eigenstates both fluctuations are of the order ${\cal O}(1)$ while for the coherent states we have normal fluctuations of $n_1$ of the order $\sqrt{n_1}$ and for the phase $\theta$ of the order of
$\frac{1}{\sqrt{n_1}}$. The later behavior is consistent with the fact that $\theta$ is an essentially classical variable -- the order parameter for the Bose-Einstein condensation of Cooper pairs.

\section{Stability of classical states}

 The reasons for a "spontaneous transition to the classical world" in the regime of large quantum numbers have been extensively discussed since the famous "Schr\"odinger Cat Paradox" \cite{joos}. Namely, the quantum states of large  systems can be observed only if they are sufficiently stable with respect to the interaction with an environment. For large enough systems the stable states are  described rather by the strongly localized symmetry-breaking solutions of the nonlinear mean-field type equations (e.g. Hartree/Hartree-Fock eqs., Born-Oppenheimer approximation) than by the eigenvectors of the exact Hamiltonians. This is for instance the case of fixed molecular structure for large enough molecules (e.g. famous chiral molecules) or deformed nuclei. For the similar reasons nonlinear mean-field Gross-Pitaevski equation should give a proper descriptions of the observable states of  $N$ Cooper pairs in contrast to the eigenvectors of  the isolated system Hamiltonians (\ref{2mod}), (\ref{2mod'}), (\ref{B-H}), (\ref{phmodq}), or (\ref{2mod1}).
\par
The complete analysis of the stability of the models given by (\ref{2mod}) and (\ref{2mod1})
will be presented in the forthcomming paper. Here we discuss briefly the model (\ref{2mod1}) assuming that the main source of instability is the transition of bosons from the condensate to
other "normal" modes and back. This effect can be modelled by adding the dissipative part to the equation of motion for the density matrix of the oscillator
\begin{equation}
\frac{d}{dt}{\hat \rho}= -i[{\hat H}_{b},{\hat \rho}] + L({\hat \rho})\ ,
\label{diss}
\end{equation}
\begin{equation}
L({\hat \rho})= \frac{1}{2}\gamma([{\hat b}{\hat \rho}, {\hat b}^{\dagger}] +[{\hat b},{\hat \rho}{\hat b}^{\dagger}] ) +\frac{1}{2}\delta([{\hat b}^{\dagger}{\hat \rho}, {\hat b}] +[{\hat b}^{\dagger},{\hat \rho}{\hat b}] )\ .
\label{diss1}
\end{equation}
A simple measure of stability of a pure state $|\phi\rangle$ is given by the initial decay rate of fidelity due to the dissipative part of the dynamics
\begin{equation}
\Gamma(|\phi\rangle)= -\langle\phi|L (|\phi\rangle\langle\phi|)|\phi\rangle \ = (\gamma +\delta)(\langle\phi|{\hat b}^{\dagger}{\hat b}|\phi\rangle -|\langle\phi|{\hat b}|\phi\rangle|^2) + \delta\ .
\label{fid}
\end{equation}
The decay rate is minimal and equal to $\delta$ for the states satisfying the eigenvector condition ${\hat b}|\phi\rangle = \alpha |\phi\rangle$ i.e. for the coherent vectors (\ref{coh1}). The Fock vectors $|k\rangle$ decay with the rates $\Gamma(|k\rangle) = (\gamma +\delta)k + \delta$ what means that the life-time of the low lying states of (\ref{2mod1}) which are supposed to support "superconducting qubit" is about $\bar n_1 \sim 10^8$ times shorter than the life-time of classical coherent states. Therefore, the states observed in experiments should be
described by mixtures of coherent vectors.
\section{Comparison with experiments}
We discuss now briefly several, already performed experimental tests aimed to justify the picture of superconducting qubits.
\par
{\it Coherent oscillations.} The charge oscillations for CPB have been confirmed for the first time in \cite{N}. The mean-field model predicts small oscillations with the plasma frequency
$\omega_0= \sqrt{2E_CE_J}$ while for the charge qubit model the frequency is a more complicated
function of $E_C, E_J$ and the applied voltage $U$ \cite{W}. It is not clear whether this difference can be tested experimentally as it seems that no independent measurements of $E_C$ and $E_J$ are available and moreover JJ's parameters are subject to time-dependent fluctuations \cite{D}.
\par
{\it Transmissivity of CPB.}  The interesting experiment \cite{Aa} shows the existence of two distinct states of CPB correlated with two values of transmissivity for CPB connected to a single electron transistor. However, classical oscillator shows also a "two state structure" described by the (averaged in time) probability distribution for its canonical variables -
$p(x)= \frac{1}{\pi}(1-x^2)^{-1/2}$ ( amplitude = 1) - with two infinite maxima at $\pm 1$.
\par
{\it Particle number fluctuations.} It seems that the large fluctuations $(\sim\sqrt{n_1})$
of the Cooper pair number predicted by the mean-field theory contradict the experimentally proved precise control of $n_1$ \cite{La}.
However, this experiment has been performed with the superconducting-normal junction. Therefore, Cooper pairs exist on the superconducting island only and the normal electrode acts as a heat baths. It follows, that in this experimental setting the state of Cooper pairs is given by the Gibbs density matrix
\begin{equation}
{\hat \rho} =  Z^{-1}\exp\Bigl\{-\frac{E_C}{kT}({\hat a}^{\dagger}_1 {\hat a}_1 -\bar{n}_1)^2\Bigr\} 
\label{gibbs}
\end{equation}
which at low temperatures $(kT < E_C)$ exhibits, indeed, very small fluctuations of ${\hat a}^{\dagger}_1 {\hat a}_1$.
\par
{\it Rabi oscillations, Ramsey fringes.} The observation of Rabi oscillations and the associated Ramsey fringes are easily and elegantly described by JJ-qubit models \cite{W}. On the other hand , as shown in \cite{G1},\cite{G2},\cite{M}, the experimental data can be attributed to classical nonlinear dynamics and statistics as well.
\par
{\it Entanglement via state tomography.} The experimental data for two coupled current-biased JJ's has been analysed in terms of the "state tomography of two phase qubits" \cite{MS}. Here again the analysis is strongly model-dependent and does not exclude the alternative interpretation in terms of coupled nonlinear oscillators.
\section{A new test of quantumness}
The presented above experimental tests, although quite convincing, are not ultimate because of their strong model-dependence.
Model independent Bell inequalities are often proposed as an ultimate test of quantumness. In the case of coupled JJ's  the corresponding experiments are still to be done. Unfortunately, even a positive result - violation of Bell inequalities - can be challenged because of the strong presence of the "locality loophole" in this setting. Namely, in contrast to experiments involving photons we are not able to separate JJ's after the act of interaction. Another interesting possibility is to test Bell inequalities for multi-time correlations in a single system as proposed in \cite{LG}.
\par
Recently, a new model-independent test of quantumness applicable to a single system has been proposed \cite{AR}. Assume that the CPB is indeed a two-level system. Then we can choose the following observables $A$ and $B$ given in a fixed basis by the following matrices:

\begin{equation}
A =\begin{pmatrix}0.724 & 0.249\cr
            0.249 & 0.0854
\end{pmatrix} \ ,
B =\begin{pmatrix} 1 & 0\cr
            0 & 0.309
\end{pmatrix}
\label{example}
\end{equation}
It is easy to check that for an arbitrary state $\psi$ the following inequalities hold
\begin{equation}
0\leq \langle \psi|A|\psi\rangle\leq \langle \psi|B|\psi\rangle .
\label{noimp}
\end{equation}
On the other hand for a specially choosen state 
$\phi = \begin{pmatrix} 0.391\cr
            0.920
\end{pmatrix}$
we obtain : $<\phi|B^2|\phi>-<\phi|A^2|\phi>=-0.0590$. This  clearly demonstrates
the quantum nature of the qubit because, for any classical model representing the observables $A,B$ by the functions on a certain "phase space" and states by the probability measures, if the inequality  $0\leq \langle A\rangle\leq \langle B\rangle$ holds for all probability measures then always $0\leq \langle A^2\rangle\leq \langle B^2\rangle$. 

\section{Conclusions}

The large body of experimental results which can be easily and elegantly interpreted in terms of superconducting qubits suggests to accept the quantumness of Josephson junctions. On the other hand, as discussed above, the idea of "macroscopic quantum system" applied to N-particle systems with N of the order of $10^8$ deeply contradicts our understanding of their quantum and statistical properties. In particular, it is generally believed and experimentally confirmed in chemistry, nuclear physics , quantum optics and physics of critical phenomena that  already for much lower particle numbers or quantum numbers, the eigenvectors of the Hamiltonians are highly unstable with respect to the interaction with an enviroment. As a consequence, the observed quantum states
are perfectly described by the stable localised symmetry breaking solutions of the non-linear mean-field type equations. These solutions satisfy classical equations of motion with additional damping terms. The system of 2N electrons in a superconducting state seems to be ideally suited for the mean-field description in terms of the Gross-Pitaevski or rather the Ginzburg-Landau equations. Therefore, it is still necessary to discuss alternative interpretations of existing experiments and search for new ones until clear evidence to quantum or classical character of JJ's becomes available. The proposal of such an ultimate experiment is discussed in this paper.

\par
\textbf{ Acknowledgements} This work is supported by the Polish Ministry
of Science and Information Technology- grant PBZ-MIN-008/P03/2003 and EC grant SCALA.

\end{document}